\documentclass[12pt,preprintnumbers,nofootinbib,letterpaper]{article}
\pdfoutput=1
\usepackage{amsmath,amssymb,amsfonts,cite,xcolor,epsf,epsfig,epstopdf,graphics,booktabs,graphicx,mathtools,subcaption,physics,verbatim}
\def\thefootnote{*\arabic{footnote}}
\definecolor{ultramarine}{rgb}{0.07, 0.04, 0.56}
\definecolor{cadmiumgreen}{rgb}{0.0, 0.42, 0.24}
\definecolor{indigo(dye)}{rgb}{0.0, 0.25, 0.42}
\usepackage[linktocpage=true,breaklinks]{hyperref}
\hypersetup{
	colorlinks=true,
	citecolor=ultramarine,
	linkcolor=cadmiumgreen,
	urlcolor=indigo(dye),
}

\usepackage[utf8]{inputenc}

\numberwithin{equation}{section}

\usepackage{array}
\newcolumntype{P}[1]{>{\centering\arraybackslash}p{#1}}

\usepackage[shortlabels]{enumitem}

\usepackage{dcolumn}
\newcolumntype{M}[1]{>{\centering\arraybackslash}m{#1}}
\newcolumntype{N}{@{}m{0pt}@{}}

\usepackage{amsmath}
\usepackage{ulem}

\usepackage[utf8]{inputenc}
\usepackage[T1]{fontenc}
\usepackage{lmodern}
\usepackage{geometry}
\usepackage{amsmath,amssymb,amsthm,mathtools}
\usepackage{bm}
\usepackage{microtype}
\usepackage{physics}

\newcommand{\Mpl}{M_{\rm Pl}}

\newcommand{\m}{m}

\newcommand{\D}{{\rm d}}

\newcommand{\be}{\begin{equation}}  
\newcommand{\ee}{\end{equation}}

\usepackage{tikz}

\begin{document}
	

\begin{center}
		
\def\thefootnote{\fnsymbol{footnote}}
		
\vspace*{1.5cm}
{\Large {\bf Imperfect dark matter with higher derivatives}}
\\[1cm]
		
{Mohammad Ali Gorji}
\\[.7cm]
		
{\small \textit{Cosmology, Gravity, and Astroparticle Physics Group, Center for Theoretical Physics of the Universe, Institute for Basic Science (IBS), Daejeon, 34126, Korea
}}
		
\end{center}
	
\vspace{1cm}
	
\hrule \vspace{0.5cm}
	
\begin{abstract}	
We introduce a higher-derivative action for dark matter whose energy-momentum tensor describes an imperfect fluid with nonzero pressure, energy flux, and anisotropic stress. In the limit where the higher-derivative couplings are switched off, the energy-momentum tensor reduces to pressureless dust. A systematic derivation follows from extending the singular conformal transformation used in the mimetic dark matter scenario to include higher-derivative terms while the resulting action is general and does not rely on the mimetic framework. On a homogeneous cosmological background, the dynamics coincides with that of pressureless dust, while in the presence of inhomogeneities the higher-derivative terms generate nonzero acceleration and vorticity, making it possible to avoid the formation of caustic singularities even if the strong energy condition satisfies. In particular, within the mimetic realization these terms can resolve the usual caustic pathology of mimetic dark matter.
\end{abstract}
	
\vspace{0.5cm} 
	
\hrule
\def\thefootnote{\arabic{footnote}}
\setcounter{footnote}{0}


	
\newpage

\section{Introduction}\label{sec:introduction}

Astrophysical and cosmological observations, such as galaxy rotation curves and the formation of large-scale structure, indicate the existence of a non-relativistic matter component beyond ordinary baryonic matter. This dominant, yet-unidentified component is known as cold dark matter. Although its origin remains unknown, the observational evidence for its presence is robust \cite{Liddle:1993fq,Bertone:2004pz,Feng:2010gw}. In the standard $\Lambda$CDM model, dark matter is modeled as pressureless dust, with energy-momentum tensor
\begin{align}\label{dust-EMT}
T^{\rm dust}_{\mu\nu} = \rho\, u_{\mu} u_{\nu} \,;
\qquad 
u^\alpha u_\alpha = -1 \,,
\end{align}
where $\rho$ is the energy density and $u^\mu$ is the four-velocity of the fluid. 

The action for a perfect fluid in terms of hydrodynamical variables was first introduced by Schutz \cite{Schutz:1970my}. For the case of irrotational dust, we have \cite{Schutz:1977df,Brown:1992kc}
\begin{align}\label{action-dust}
S_{\rm dust} = -\int \D^4x\sqrt{-g} \lambda \left(u^\alpha{u}_\alpha+1\right) \,;
\qquad
\rho = 2\lambda \,,
\end{align}
where $g_{\mu\nu}$ is the spacetime metric, $u_\mu$ is a hypersurface-orthogonal four-velocity, and we work with metric signature $(-,+,+,+)$. Taking variation with respect to the auxiliary field $\lambda$ or equivalently energy density $\rho$ gives the normalization condition for the four-velocity while taking variation with respect to the metric correctly reproduces the dust energy-momentum tensor \eqref{dust-EMT} that is defined as $T^{\rm dust}_{\mu\nu}=-(2/\sqrt{-g})\delta{S}_{\rm dust}/\delta{g}^{\mu\nu}$. 

Remarkably, the dust action \eqref{action-dust} can be obtained by isolating the conformal mode of the metric within the mimetic dark matter scenario \cite{Chamseddine:2013kea}. Consider the singular conformal transformation \cite{Chamseddine:2013kea,Barvinsky:2013mea}
\begin{align}\label{trans-mimetic}
g_{\mu\nu} = - \left( \tilde{g}^{\alpha\beta} u_\alpha u_\beta \right) \tilde{g}_{\mu\nu} \,,
\end{align}
where $\tilde{g}_{\mu\nu}$ is an auxiliary metric. The inverse contravariant metric is given by $g^{\mu\nu} = - \left( \tilde{g}^{\alpha\beta} u_\alpha u_\beta \right)^{-1} \tilde{g}^{\mu\nu}$ which after contracting both sides with $u_\mu{u}_\nu$ gives $u^\alpha u_{\alpha} = -1$. Note that we cannot express auxiliary metric $\tilde{g}_{\mu\nu}$ in terms of the physical metric $g_{\mu\nu}$. This shows that the conformal transformation \eqref{trans-mimetic} is singular and, therefore, imposing it at the level of the action yields a dynamically inequivalent theory. We discuss this in detail in the next section. Substituting \eqref{trans-mimetic} directly into the Einstein-Hilbert action\footnote{We work in units $\hbar=c=1$ such that $\Mpl=1/\sqrt{8\pi{G}}$ is the reduced Planck mass.} $S_{\rm E.H.} = \tfrac{\Mpl^2}{2} \int\D^4x \sqrt{-g} R$ yields field equations equivalent to general relativity sourced by pressureless dust; the dust arises from the isolated conformal degree of freedom of the metric. Subsequently, it was shown that applying the singular transformation \eqref{trans-mimetic} to the Einstein-Hilbert action is classically equivalent to working entirely with the physical metric and supplementing the Einstein-Hilbert action with the dust action \eqref{action-dust} \cite{Golovnev:2013jxa}.

Although the Schutz formulation of dust \eqref{dust-EMT} and the mimetic dark matter action are equivalent, in the mimetic construction the dust-like component arises from the conformal degree of freedom of the metric. For pressureless dust, conservation of the energy–momentum tensor, $\nabla_\alpha{T}^\alpha{}_\mu=0$, implies that the four-velocity satisfies the geodesic equation
\begin{align}\label{geodesic}
u^\alpha \nabla_\alpha u^\mu = 0 \,.
\end{align}
Because congruences of geodesics generally focus, dust flows develop caustics (shell crossing) in finite time. When the dust is identified with the conformal mode, such caustics manifest as singular behavior of the mimetic field, signaling a breakdown of the mimetic reconstruction of the metric \cite{Gorji:2020ten}. To prevent caustic formation, non-vanishing acceleration (deviation from the geodesics equation \eqref{geodesic}) or non-vanishing vorticity is necessary. For example, one may add $F_{\mu\nu}=2\partial_{[\mu} u_{\nu]}$ to give dynamics to the transverse modes leading to non-vanishing vorticity which might avoid formation of caustics \cite{Barvinsky:2013mea,Chaichian:2014qba,Domenech:2025qny}.\footnote{One may also consider gauge-field extensions of the mimetic construction, which are generally free of caustics pathology \cite{Gorji:2018okn,Jirousek:2018ago,Gorji:2019ttx,Hammer:2020dqp,Colleaux:2025vtm}. However, in those cases the mimetic sector no longer reproduces dark matter phenomenology.} In this paper, we instead focus on the role of higher-derivative couplings. In order to do so, we first systematically take into account higher-derivative terms within the mimetic framework, and then show that higher-derivative terms generally induce both acceleration and vorticity, thereby regulating the flow allowing caustics to be avoided.

The rest of the paper is organized as follows. In Sec.~\ref{sec-HD-DT} we present a systematic generalization of the mimetic construction that incorporates higher-derivative terms. In Sec.~\ref{sec-HD-mimetic} we introduce a new higher-derivative dark matter action and we derive the corresponding equations of motion. In Sec.~\ref{sec-ADM}, using a $3+1$ decomposition, we show that on homogeneous cosmological backgrounds the theory behaves as pressureless dark matter even in the presence of higher-derivative corrections. In Sec.~\ref{sec-caustic} we demonstrate that higher-derivative terms can generate repulsive inhomogeneous forces and vorticity, allowing caustic formation to be avoided. Sec.~\ref{sec:summary} is devoted to the summary and conclusions. Additional technical computations are presented in Appendix~\ref{app:EMT-eigenvalue}.

\section{Singular disformal transformations}\label{sec-HD-DT}

The conformal transformation \eqref{trans-mimetic} is singular: the map $\tilde g_{\mu\nu}\!\to\! g_{\mu\nu}$ is non-invertible, so $\tilde g_{\mu\nu}$ cannot be expressed in terms of $g_{\mu\nu}$. Our goal is to construct a higher-derivative generalization of \eqref{trans-mimetic}. To this end, we first recall that \eqref{trans-mimetic} arises as a non-invertible (singular) limit of a more general disformal transformation~\cite{Deruelle:2014zza,Mirzagholi:2014ifa,Arroja:2015wpa,Domenech:2015tca,BenAchour:2024hbg}. This observation provides a systematic framework for constructing higher-derivative extensions of \eqref{trans-mimetic}. Readers primarily interested in the final result, rather than the derivation, may proceed directly to the next section and start from the action \eqref{action}.

The disformal transformation for scalar-tensor theories is given by
\begin{align}\label{eq:DT}
g_{\mu\nu} = A(\phi,\tilde{X}) \tilde{g}_{\mu\nu} + B(\phi,\tilde{X}) \phi_\mu \phi_\nu \,;
\qquad
\tilde{X} \equiv \tilde{g}^{\alpha\beta}\phi_\alpha\phi_\beta \,,
\end{align}
where we have used the notation $\phi_\mu=\nabla_\mu\phi$ such that $\phi_{\mu\nu}=\nabla_\nu\nabla_\mu\phi$ and so on. The inverse contravariant metric is given by
\begin{align}
g^{\mu\nu} = \frac{1}{A} \tilde{g}^{\mu\nu} - \frac{B}{A(A+{\tilde X}B)} \tilde{\phi}^\mu \tilde{\phi}^\nu \,;
\qquad
\tilde{\phi}^{\mu} \equiv \tilde{g}^{\mu\alpha} \phi_\alpha \,.
\end{align}
We adopt the convention that a tilde on any quantity indicates that all metrics appearing in it are the auxiliary metric $\tilde{g}_{\mu\nu}$. 

To find the singular limit, we look at the eigenvalue problem \cite{Zumalacarregui:2013pma,Jirousek:2022rym,Jirousek:2022jhh}
\begin{align}\label{eq:eigen-Eq}
J^{\alpha\beta}_{\mu\nu} \xi^{a}_{\alpha\beta} = \lambda^{a} \xi^{a}_{\mu\nu} \,,
\qquad
J^{\mu\nu}_{\alpha\beta} \zeta_{a}^{\alpha\beta} = \lambda^{a} \zeta_{a}^{\mu\nu} \,,
\end{align}
where $\xi^{a}_{\mu\nu}$ is the eigentensor and $\zeta_{a}^{\mu\nu}$ is the dual eigentensor, both correspond to the eigenvalues $\lambda^{a}$ and 
\begin{align}
J^{\rho\sigma}_{\mu\nu} \equiv \frac{\partial{g}_{\mu\nu}}{\partial{\tilde{g}}^{\rho\sigma}} \,,
\end{align}
is the Jacobian of the transformation. For the disformal transformation \eqref{eq:DT}, the Jacobian takes the form \cite{Zumalacarregui:2013pma}
\begin{align}\label{eq:J-DT}
J^{\rho\sigma}_{\mu\nu} = 
A\, \delta^{(\rho}_\mu \delta^{\sigma)}_\nu 
- \left( A_{\tilde{X}} \tilde{g}_{\mu\nu} + B_{\tilde{X}} \phi_{\mu} \phi_{\nu} \right) \tilde{\phi}^{\rho} \tilde{\phi}^{\sigma} \,.
\end{align}
Solving Eqs. \eqref{eq:eigen-Eq} for the above Jacobian, we find the following eigenvalues
\begin{align}\label{eq:eigenvalues-DT}
\lambda_0 = A
\,,
\qquad
\lambda_\star = A - A_{\tilde{X}} \tilde{X} - B_{\tilde{X}} \tilde{X}^2
\,.
\end{align}
Substituting $\lambda_\star$ in \eqref{eq:eigen-Eq}, we find $\xi^\star_{\mu\nu} = A_{\tilde{X}} \tilde{g}_{\mu\nu} + B_{\tilde{X}} \phi_{\mu} \phi_{\nu}$ and 
$\zeta_\star^{\mu\nu} = \tilde{\phi}^{\mu} \tilde{\phi}^{\nu}$
while for $\lambda_0$ we find $\xi^0_{\mu\nu}$ and $\zeta_0^{\mu\nu}$ which satisfy $\xi^0_{\mu\nu}\zeta_\star^{\mu\nu}=0$ and $\zeta_0^{\mu\nu}\xi^\star_{\mu\nu}=0$. Now, we look at the singular limits. The case $\lambda_0=0$ is not a valid choice as there will be no inverse contravariant metric. We thus focus on 
\begin{align}\label{eq:lambda-0-S}
\lambda_\star = 0 \,
\quad
\Rightarrow
\quad
B = - \frac{A}{{\tilde{X}}} - \frac{1}{\epsilon(\phi)} \,,
\end{align}
where $\epsilon(\phi)\neq0$ is an arbitrary function which can be normalized as $\epsilon=\pm1$ by redefinition of $\phi$. We then have
\begin{align}\label{eq:X-S}
X=\frac{{\tilde{X}}}{A+B{\tilde{X}}} 
\,,
\end{align}
where $X\equiv{g}^{\alpha\beta}\phi_\alpha\phi_\beta$. Substituting $B$ given by Eq. \eqref{eq:lambda-0-S} in \eqref{eq:X-S} leads to the constraint
\begin{align}\label{eq:mimetic-constraint}
X=-1 \,,
\end{align}
where we have set $\epsilon=1$ to impose that $\phi^\mu$ is timelike. Note that this constraint is written completely in terms of the physical metric $g_{\mu\nu}$. The original mimetic scenario deals with the conformal case $B=0$. Then, Eq.~\eqref{eq:lambda-0-S} gives
\begin{align}
 A=-{\tilde X} \,,
\end{align}
and we find the singular conformal transformation \eqref{trans-mimetic}. 

As noted above, applying \eqref{trans-mimetic} to the Einstein–Hilbert action is classically equivalent to working entirely with the physical metric and supplementing it with the dust action \eqref{action-dust}~\cite{Golovnev:2013jxa}. We therefore look for a higher-derivative extension of the condition $A=-\tilde X$. However, extending the disformal map \eqref{eq:DT} to include second (or higher) derivatives of the scalar is subtle~\cite{Ezquiaga:2017ner,Babichev:2019twf,Babichev:2021bim,Babichev:2024eoh}: a straightforward modification that adds a term proportional to $\phi_{\mu\nu}$ renders the transformation non-invertible~\cite{Babichev:2021bim}. Instead, viable higher-derivative generalizations require more structured dependence on $X_\mu=\nabla_\mu{X}$, for which invertible constructions are known. As shown in Ref.~\cite{Takahashi:2021ttd}, second derivatives of the scalar field can be incorporated consistently via
\begin{align}\label{eq:DT-HD}
g_{\mu\nu}
= A\,\tilde g_{\mu\nu}
+ B\,\phi_\mu \phi_\nu
+ 2C\,\phi_{(\mu}\tilde X_{\nu)}
+ D\,\tilde X_\mu \tilde X_\nu \,,
\end{align}
where $\tilde\phi^\mu\!\equiv\!\tilde g^{\mu\alpha}\phi_\alpha$, $\tilde X\!\equiv\!\tilde g^{\alpha\beta}\phi_\alpha\phi_\beta$, $\tilde X_\mu\!\equiv\!\tilde\nabla_\mu \tilde X$, and $\phi_{(\mu}\tilde X_{\nu)}\!\equiv\!(\phi_\mu\tilde X_\nu+\phi_\nu\tilde X_\mu)/2$ and the coefficients $A,B,C,D$ are functions of the scalar field $\phi$ and of three independent scalars,
\begin{align}\label{XYZ-tilde-def}
\tilde X \equiv \tilde g^{\alpha\beta}\phi_\alpha\phi_\beta\,,
\qquad
\tilde Y \equiv \tilde g^{\alpha\beta}\phi_\alpha \tilde X_\beta\,,
\qquad
\tilde Z \equiv \tilde g^{\alpha\beta}\tilde X_\alpha \tilde X_\beta \,.
\end{align}
One can easily verifies that the ansatz \eqref{eq:DT-HD} is closed under composition (matrix multiplication), so the inverse metric has the same structure,
\begin{align}\label{eq:DT-HD-inv}
g^{\mu\nu}
= \bar A\,\tilde g^{\mu\nu}
+ \bar B\,\tilde\phi^\mu \tilde\phi^\nu
+ 2\bar C\,\tilde\phi^{(\mu}\tilde X^{\nu)}
+ \bar D\,\tilde X^{\mu}\tilde X^{\nu}\,,
\end{align}
where $\tilde X^\mu\!\equiv\!\tilde g^{\mu\nu}\tilde X_\nu$, and the explicit expressions for $\bar A,\bar B,\bar C,\bar D$ are given in Ref.~\cite{Takahashi:2021ttd}. Note that even if one sets $D=0$ ($C=0$) in \eqref{eq:DT-HD}, the inverse metric \eqref{eq:DT-HD-inv} generally exhibits ${\bar D}\neq 0$ (${\bar C}\neq 0$). Hence, once the $C$-term ($D$-term) is included, consistency requires allowing a nonzero $D$-term ($C$-term) as well. Thus, we need to take into account both of them together. 

The disformal transformation \eqref{eq:DT-HD} includes terms up to second derivatives of the scalar field $\phi_{\mu\nu}$ in a way that ensures the existence of the contravariant inverse metric \eqref{eq:DT-HD-inv}, built from the same elements \eqref{XYZ-tilde-def}. At leading order, with only first derivatives $\phi_\alpha$, the only building block is $\tilde X$ and the disformal transformation takes the well-known form \eqref{eq:DT}. At the next order, with second derivatives $\phi_{\mu\nu}$, consistency requires both $\tilde Y$ and $\tilde Z$ and we find \eqref{eq:DT-HD}. Extending to third derivatives $\phi_{\mu\nu\rho}$, seven additional building blocks are needed \cite{Takahashi:2021ttd}
\begin{align}\label{third-derivative-quantities}
\phi_\alpha \tilde{Y}^\alpha , \,
\phi_\alpha \tilde{Z}^\alpha, \,
\tilde{X}_\alpha \tilde{Y}^\alpha, \,
\tilde{X}_\alpha \tilde{Z}^\alpha, \,
\tilde{Y}_\alpha \tilde{Y}^\alpha, \,
\tilde{Y}_\alpha \tilde{Z}^\alpha, \,
\tilde{Z}_\alpha \tilde{Z}^\alpha  \,.
\end{align}
It is important to note that although $\phi_\alpha \tilde{Y}^\alpha$ contains up to second derivatives of the scalar field, it did not show up in the systematic construction of the disformal transformation at second order. This shows that how the consistency of the disformal transformation restricts the situation: $\phi_\alpha \tilde{Y}^\alpha$ should be taken into account together with all other six terms in \eqref{third-derivative-quantities} to construct a consistent disformal transformation up to the third derivatives of the scalar field. This convention preserves closure under matrix multiplication and ensures that an inverse metric with the same structure exists, keeping the transformation invertible.

Having confirmed the consistency of the higher-derivative disformal transformation \eqref{eq:DT-HD}, we look at the conformal subset to find the higher-derivative extension of the mimetic gravity. In this case, the Jacobian takes the form
\begin{align}\label{eq:J-DT-HD}
J^{\rho\sigma}_{\mu\nu} = 
A\, \delta^{(\rho}_\mu \delta^{\sigma)}_\nu 
- \left[ A_{\tilde{X}} \tilde{\phi}^{\rho} \tilde{\phi}^{\sigma} 
+ 2 A_{\tilde{Y}} \tilde{\phi}^{(\rho} \tilde{X}^{\sigma)}
+ A_{\tilde{Z}} \tilde{X}^{\rho} \tilde{X}^{\sigma}
+ 2 A_{\tilde{Z}} \tilde{X}^{\alpha} \tilde{\nabla}_\alpha\left(
\tilde{\phi}^\rho \tilde{\phi}^\sigma \right)
\right] \tilde{g}_{\mu\nu} + \cdots \,,
\end{align}
where the ellipses denote contributions involving the third-order terms listed in Eq.~\eqref{third-derivative-quantities}. 

Solving Eqs. \eqref{eq:eigen-Eq} for the above Jacobian with $\xi^\star_{\mu\nu} \propto \tilde{g}_{\mu\nu}$, we find the following eigenvalues
\begin{align}\label{eq:eigenvalues-DT-HD}
\lambda_0 = A
\,,
\qquad \lambda_\star 
= A - A_{\tilde{X}} \tilde{X} - 2 A_{\tilde{Y}} \tilde{Y} - 3 A_{\tilde{Z}} \tilde{Z}
\,.
\end{align}
Similar to the above analysis, we look at the singular limit
\begin{align}\label{eq:lambda-0-S-HD}
\lambda_\star = 0 \,
\quad
\Rightarrow
\quad
A = \tilde{X} f\left(\phi,\tfrac{\tilde{Y}}{\tilde{X}^2},\tfrac{\tilde{Z}}{\tilde{X}^3}\right) \,,
\end{align}
where $f$ is a general analytical function.

To leading order in a derivative expansion, one finds
\begin{align}\label{XYZ-trans}
X=\frac{\tilde X}{A} \,,
\qquad
Y=\frac{\tilde Y}{A^{2}}+\cdots \,,
\qquad
Z=\frac{\tilde Z}{A^{3}}+\cdots \,,
\end{align}
where 
\begin{align}\label{XYZ-def}
X \equiv g^{\alpha\beta}\phi_\alpha\phi_\beta \,,
\qquad
Y \equiv g^{\alpha\beta}\phi_\alpha X_\beta \,,
\qquad
Z \equiv g^{\alpha\beta}X_\alpha X_\beta \,.
\end{align}
The ellipses denote contributions involving derivatives of $A$ which include the third derivatives of $\phi$. As explained above, a consistent treatment of these third derivatives requires adding the seven extra building blocks \eqref{third-derivative-quantities}, which modify the result \eqref{eq:lambda-0-S-HD}.

Note that, similar to the usual mimetic transformation \eqref{trans-mimetic}, the generalized higher-derivative conformal transformation $g_{\mu\nu}=A\,\tilde g_{\mu\nu}$ with $A$ given by \eqref{eq:lambda-0-S-HD} is invariant, at the order retained in this section, under a local Weyl rescaling of the auxiliary metric $\tilde g_{\mu\nu}\to \Omega^{2}(x)\tilde g_{\mu\nu}$: indeed, $\tilde X\to \Omega^{-2}\tilde X$ while the dimensionless ratios $\tilde Y/\tilde X^{2}$ and $\tilde Z/\tilde X^{3}$ are invariant at this order, so that $A\to \Omega^{-2}A$ and consequently $g_{\mu\nu}=A\,\tilde g_{\mu\nu}$ remains unchanged. This auxiliary Weyl redundancy is the hallmark of the non-invertibility of the map within our approximation and, as in standard mimetic gravity, it is precisely what isolates the conformal mode.

We now rewrite the result \eqref{eq:lambda-0-S-HD} in terms of quantities built from the physical metric. Using \eqref{XYZ-trans} in \eqref{eq:lambda-0-S-HD}, we obtain
\begin{align}\label{eq:mimetic-constraint-HD}
X\,f\!\left(\phi,\tfrac{Y}{X^{2}},\tfrac{Z}{X^{3}}\right)=1\,.
\end{align}
This constraint is expressed purely in terms of the physical metric $g_{\mu\nu}$ and generalizes the mimetic constraint \eqref{eq:mimetic-constraint} to include second derivatives of the scalar field. In the absence of higher-derivative terms, $f$ becomes only function of $\phi$ which can be renormalized to $f=\pm1$. For the case of timelike $\phi^\mu$, we need to choose $f=-1$ and \eqref{eq:mimetic-constraint-HD} reduces to \eqref{eq:mimetic-constraint}. As a simple illustrative choice, one may take
$f=c_1(\phi) + c_2(\phi)  |Y|^{1/2}/X + c_3(\phi)  |Z|^{1/3}/X$ such that the mimetic constraint is $c_1(\phi)  X + c_2(\phi)  |Y|^{1/2} + c_3(\phi)  |Z|^{1/3}=1$. In what follows we keep $f$ arbitrary.

It is worth noting that, at first sight, the above setup may resemble two-field extensions of the mimetic scenario \cite{Firouzjahi:2018xob,Shen:2019nyp,Mansoori:2021fjd,Zheng:2022vwm}. However, the present construction remains effectively single-field: $X=g^{\alpha\beta}\phi_\alpha\phi_\beta$ (and thus $Y$ and $Z$) is a composite built from the same scalar $\phi$ and the metric, rather than an independent field. Consequently, the structures and constraints we employ here do not straightforwardly map onto those of two-field mimetic models.

\section{Higher-derivative mimetic gravity}\label{sec-HD-mimetic}
As shown in the previous section, systematically including second-derivative terms modifies only the standard mimetic constraint \eqref{eq:mimetic-constraint}, replacing it by \eqref{eq:mimetic-constraint-HD}. The action for the corresponding higher-derivative mimetic gravity therefore is
\begin{align}\label{action}
\boxed{
S=\int \D^4x\,\sqrt{-g}\,\bigg[
\frac{M_{\rm Pl}^2}{2}\,R
+\lambda\!\left(X\,f\!\left(\tfrac{Y}{X^2},\tfrac{Z}{X^3}\right)-1\right)
\bigg] \,,
}
\end{align}
where $f$ is a free function and $X,Y,Z$ are defined in \eqref{XYZ-def}. For concreteness we have imposed the shift symmetry so $f$ depends only on the dimensionless ratios $Y/X^2$ and $Z/X^3$. In the limit $f\to{-1}$, one recovers the original mimetic action \cite{Golovnev:2013jxa}.

Taking variation with respect to auxiliary field $\lambda$, we find the generalized mimetic constraint
\begin{align}\label{mimetic-constraint}
Xf\left(\tfrac{Y}{X^2},\tfrac{Z}{X^3}\right)= 1 \,. 
\end{align}

Varying the action \eqref{action} with respect to the metric yields
\begin{align}
\Mpl^2 G_{\mu\nu} = T_{\mu\nu} \,,
\end{align}
where $G_{\mu\nu}=R_{\mu\nu}-g_{\mu\nu}R/2$ and $T_{\mu\nu}$
\begin{align}\label{Eq:EMT}
T_{\mu\nu} = - 2\lambda \left( C_{\phi\phi}\, \phi_\mu \phi_\nu + C_{\phi{X}}\, \phi_{(\mu} X_{\nu)} + C_{XX}\, X_\mu X_\nu \right) \,,
\end{align}
where we have used the mimetic constraint \eqref{mimetic-constraint} to get rid of the term $ \lambda \left(Xf -1\right) g_{\mu\nu}$. Here
\begin{align}\label{Cs-def}
\begin{split}
C_{\phi\phi} \equiv \left(Xf\right)_{,X} 
- \frac{1}{\lambda} \nabla_{\alpha} \left(\lambda C_{\phi{X}} \phi^\alpha
+ 2 \lambda C_{XX} X^\alpha
\right) \,,
\quad
C_{\phi{X}} \equiv 
\left(Xf\right)_{,Y}  \,,
\quad
C_{XX} \equiv 
\left(Xf\right)_{,Z}  \,.
\end{split}
\end{align}

Finally variation of the action \eqref{action} with respect to the scalar field gives
\begin{align}
\nabla_\alpha J^\alpha = 0 \,;
\qquad
J^\mu = \lambda \left( 
2 C_{\phi\phi} \phi^\mu
+ C_{\phi{X}} X^\mu
\right) \,,
\end{align}
where $J^\mu$ is the current for the shift symmetry. We can check that $\nabla_\alpha \left[ T^{\alpha\mu} + \lambda \left(Xf-1\right)g^{\alpha\mu}\right] = - (\nabla_\alpha J^\alpha) \phi^\mu$ as it is implied by the diffeomorphism invariance of the theory.

The action \eqref{action} contains up to second derivatives of $\phi$ (through $X_\mu$ and $Z$) and the corresponding equations of motion involve up to fourth-order derivatives of $\phi$. Such higher-order equations typically signal an Ostrogradsky instability~\cite{Woodard:2015zca}. However, the situation changes in the mimetic scenario since the system is degenerate \cite{Chaichian:2014qba,Malaeb:2014vua,Takahashi:2017pje,BenAchour:2017ivq,Zheng:2018cuc,Ganz:2018mqi,Shen:2019nyp,deCesare:2020got}. In our construction the higher-derivative dependence appears only through the constraint $X\,f(Y/X^{2},Z/X^{3})=1$ enforced by the equation of the auxiliary field $\lambda$. This imposes a strong (algebraic) relation among $X$, $Y=\!\phi^\alpha X_\alpha$, and $Z=\!X^\alpha X_\alpha$, i.e. among second derivatives of $\phi$ and its first derivatives. As a result, after using the constraint $X f=1$ in the field equations, all higher-derivative contributions can be reduced to the lower-derivative quantities. This might (or might not) be enough to eliminate the would-be Ostrogradsky ghost. Therefore, a complete Hamiltonian analysis would provide a rigorous confirmation of the constraint structure for \eqref{action}, and we leave this for future work. 

On the other hand, there is a particularly simple subclass of the theory in which the potentially dangerous higher-time-derivative structure is absent, so that the would-be Ostrogradsky mode is eliminated already at the level of the operator basis. This is obtained by restricting the constraint function to depend only on the following combination
\begin{equation}\label{C-def}
{\cal C} \;\equiv\; \frac{Z}{X^3}-\left(\frac{Y}{X^2}\right)^2 \;.
\end{equation}
Introducing the transverse spatial vector
\begin{equation}\label{V-def}
V_\mu \;\equiv\; X_\mu-\frac{Y}{X}\phi_\mu,
\qquad
\phi^\mu V_\mu=0,
\end{equation}
one immediately finds
\begin{equation}
{\cal C}=\frac{V_\mu V^\mu}{X^3}\,.
\end{equation}
In this form, it is clear that ${\cal C}$ is constructed solely from the component of $X_\mu$ orthogonal to $\phi_\mu$. In particular, it does not probe the ``longitudinal” component along $\phi_\mu$, which is responsible for higher time-derivative contributions. This is especially transparent in unitary gauge, $\phi=t$. Using the ADM decomposition ${\D}s^2 = - N^2 {\D}t^2 + h_{ij}({\D}x^i + N^i {\D}t)({\D}x^j+N^j {\D}t)$, one finds $X=-1/N^2$ and ${\cal C}=h^{ij}\partial_i X\partial_j X/X^3$, which involves no time derivative of the lapse (and hence no second time derivatives of the scalar field). The situation is therefore closely analogous to the standard mimetic case: the mimetic constraint reduces to 
\begin{align}
Xf\left({\cal C}\right)=1 \,,
\end{align}
and, in unitary gauge, it is a first-order equation for the scalar field. Accordingly, the number of propagating degrees of freedom in unitary gauge matches the usual mimetic scenario: two tensor modes and one scalar mode.

It is worth noting that the combination $\eqref{C-def}$ is not contained in Horndeski or, more generally, in Degenerate Higher-Order Scalar-Tensor (DHOST) theories \cite{Langlois:2015cwa}, but instead arises in the context of U-DHOST theories \cite{DeFelice:2018ewo,DeFelice:2021hps,DeFelice:2022xvq}. In U-DHOST, the theory propagates three degrees of freedom (two tensors and one scalar) in unitary gauge, while beyond unitary gauge an additional mode appears; however, this extra mode is instantaneous, with an elliptic equation of motion, and is therefore harmless. Moreover, the ``scordatura” mechanism, which provides higher spatial derivatives while keeping higher time derivatives suppressed below a cutoff \cite{Motohashi:2019ymr,Gorji:2020bfl,Gorji:2021isn}, is built in automatically in U-DHOST. Consequently, the subset of our higher-derivative setup restricted to $\eqref{C-def}$ can also yield healthy higher-spatial-derivative behavior in the high-energy regime where higher-derivative operators become relevant.

The covariant equations of motion significantly simplify once we restrict our setup to $f({\cal C})$: the coefficients in \eqref{Cs-def} simplify to
\begin{equation}\label{new-coeff-def}
C_{\phi X}= -\frac{2Y}{X}\,C_{XX} = -\frac{2Y}{X^3}\,f_{\cal C} \,,
\end{equation}
and the energy-momentum tensor \eqref{Eq:EMT} can be written in the compact form
\begin{equation}\label{eq:EMT-fC}
	T_{\mu\nu}
	=-2\lambda\Big[
	\widehat C_{\phi\phi}\,\phi_\mu\phi_\nu
	+ \frac{f_{\cal C}}{X^2}\,V_\mu V_\nu
	\Big],
\end{equation}
where
\begin{equation}\label{eq:Chatphiphi-final}
	\widehat C_{\phi\phi}\equiv C_{\phi\phi}-\frac{Y^2}{X^2}C_{XX}
	=
	f-3{\cal C} f_{\cal C}
	-\frac{2}{\lambda}\nabla_\alpha\!\left(\lambda\,\frac{f_{\cal C}}{X^2}\,V^\alpha\right).
\end{equation}
The result \eqref{eq:EMT-fC} shows that $\phi_\mu$ is the unique time-like direction and can therefore be used as the fluid four-velocity, $u_\mu\propto\phi_\mu$. The remaining vector $V_\mu$ is purely spatial and thus provides the only additional direction available to construct imperfect contributions, which are entirely encoded in the single tensor structure $V_\mu V_\nu$. We will discuss the associated fluid interpretation and its consequences in more detail in the next section.

Moreover, the fact that in our construction all higher-derivative effects enter exclusively through the (modified) constraint sector in the action \eqref{action} makes it qualitatively different from the standard ``higher-derivative mimetic” setups, where one keeps the original mimetic constraint $X=-1$ (implemented via a term $\lambda(X+1)$) and adds higher-derivative operators directly in the Lagrangian, e.g. through functions of $\Box\phi$ and/or other higher-derivative combinations~\cite{Chamseddine:2014vna,Mirzagholi:2014ifa,Capela:2014xta,Arroja:2015wpa,Arroja:2015yvd,Cognola:2016gjy,BenAchour:2016cay,Babichev:2016jzg,Chamseddine:2016uef,Chamseddine:2016ktu,Firouzjahi:2017txv,Langlois:2017hdf,Hirano:2017zox,Zheng:2017qfs,Takahashi:2017pje,Gorji:2017cai,Langlois:2018jdg,HosseiniMansoori:2020mxj,Domenech:2023ryc}. In those approaches the constraint is fixed to $X\equiv\phi_\mu\phi^\mu=-1$, so that $\nabla_\mu X=2\phi^\alpha\phi_{\alpha\mu}=0$ identically. In particular, for the commonly used higher-derivative scalars $Y\equiv2\phi^\alpha\phi^\beta\phi_{\alpha\beta}$ and $Z\equiv4\phi^\alpha\phi^\beta\phi_{\alpha\eta}\phi_\beta{}^{\eta}$ one has $Y=0$ and $Z=0$ on the constraint surface, and similarly the “longitudinal” building blocks $\vartheta_n\equiv\phi^\alpha\phi^\beta(\phi_{\alpha\beta}^n)$ vanish for $n\geq1$. This observation was emphasized in the DHOST extension of mimetic gravity~\cite{Langlois:2018jdg}, where, precisely because $\vartheta_n=0$ when $X=-1$, the nontrivial higher-derivative operators may be chosen to depend instead on the trace-type combinations $\chi_n\equiv g^{\alpha\beta}(\phi_{\alpha\beta}^n)$ (in particular $\chi_1=\Box\phi$). By contrast, in our theory the constraint itself is deformed (schematically of the form $Xf=1$), so $X$ is not fixed to a constant and $\nabla_\mu X$ need not vanish. As a result, $Y$ and $Z$ (equivalently $\vartheta_1$ and $\vartheta_2$) are generically nonzero and become natural, non-redundant carriers of higher-derivative effects; this is in fact what one finds by a systematic analysis of the relevant singular limit, which indicates that the deformation of the constraint is essential and that the appropriate higher-derivative information is encoded in the ``longitudinal” structures rather than being pushed into $\Box\phi$-type terms.

As an illustration of the physical distinction, if the constraint is left unchanged so that $\phi^\alpha\phi_\alpha=-1$, then $X_\mu=2\phi^\alpha\phi_{\alpha\mu}=0$ and the energy-momentum tensor reduces to that of dust with $\rho=2\lambda$ and $u^\mu=\phi^\mu$. In the usual mimetic case $u_\mu=\partial_\mu\phi$ is hypersurface-orthogonal (hence irrotational) and the flow is geodesic, so caustics generically form. By contrast, in our model \eqref{action} the constraint is modified; $u^\mu$ is no longer aligned with $\partial^\mu\phi$ and need not be hypersurface-orthogonal, so the higher-derivative contributions generically induce acceleration and vorticity, allowing caustic formation to be avoided. A detailed analysis of caustics will be presented in Sec.~\ref{sec-caustic}.

\section{$1+3$ decomposition of energy-momentum tensor}\label{sec-ADM}
Unlike the standard mimetic case, where the effective energy–momentum tensor has the dust form with $\rho=2\lambda$ and $u_\mu=\partial_\mu\phi$, the tensor in \eqref{Eq:EMT} contains additional structures $\phi_{(\mu}X_{\nu)}$ and $X_\mu X_\nu$ and therefore describes an imperfect fluid. To characterize its properties, we perform a $3\!+\!1$ decomposition.

For an observer with four-velocity $u^\mu$, which is normalized as $u_\alpha u^\alpha=-1$, we define the projection tensor
\begin{align}
h_{\mu\nu} \equiv g_{\mu\nu} + u_\mu u_\nu \,,
\end{align}
so that $h^\mu{}_\alpha u^\alpha=0$ and $h^\mu{}_\alpha h^\alpha{}_\nu = h^\mu{}_\nu$. The energy–momentum tensor then decomposes as
\begin{align}\label{EMT-dec}
\begin{split}
T^{\mu\nu} &= \rho\,u^\mu u^\nu + 2u^{(\mu} q^{\nu)} + \tau^{\mu\nu}
\\
&= \rho\,u^\mu u^\nu + 2u^{(\mu} q^{\nu)} + p\,h^{\mu\nu} + \pi^{\mu\nu},
\end{split}
\end{align}
where
\begin{align}
\rho \equiv T^{\alpha\beta}u_\alpha u_\beta \,,
\qquad
q^\mu \equiv -\,h^\mu{}_\alpha\, T^{\alpha\beta} u_\beta \,,
\qquad
\tau^{\mu\nu} \equiv h^\mu{}_\alpha h^\nu{}_\beta\, T^{\alpha\beta} \,,
\end{align}
are the energy density, energy flux (momentum density in the $u^\mu$-frame), and spatial stress. By construction,
\begin{align}
q^\alpha u_\alpha = 0 \,,
\qquad \tau^{\mu\alpha} u_\alpha = 0 \,.
\end{align}
The trace and traceless parts of $\tau^{\mu\nu}$ define the isotropic pressure and the anisotropic stress
\begin{align}
p \equiv \tfrac{1}{3} h_{\alpha\beta}\tau^{\alpha\beta},\qquad
\pi^{\mu\nu} \equiv \tau^{\mu\nu} - \tfrac{1}{3} (h_{\alpha\beta}\tau^{\alpha\beta})\, h^{\mu\nu},
\end{align}
with $\pi^\alpha{}_\alpha=0$ and $\pi^{\mu\alpha}u_\alpha=0$.

We also decompose the vectors $\phi^\mu$ and $X^\mu$ into parts parallel and perpendicular to $u^\mu$ as
\begin{align}\label{phi-X-decomposition}
\begin{split}
\phi^\mu &= -\varphi_\perp\,u^\mu + \varphi^\mu \,;
\qquad 
\varphi_\perp\equiv u_\alpha\phi^\alpha \,,
\quad
\varphi^\mu\equiv h^\mu{}_\alpha \phi^\alpha \,,
\\
X^\mu &= -\chi_\perp\,u^\mu + \chi^\mu \,;
\qquad 
\chi_\perp\equiv u_\alpha X^\alpha \,,
\quad
\chi^\mu\equiv h^\mu{}_\alpha X^\alpha \,,
\end{split}
\end{align}
so that $u_\alpha\varphi^\alpha=u_\alpha\chi^\alpha=0$. 

Using the energy-momentum tensor \eqref{Eq:EMT} and the above definitions, the fluid variables are
\begin{align}\label{rho-general}
\rho &= -2\lambda \big(C_{\phi\phi}\,\varphi_\perp^2 + C_{\phi X}\,\varphi_\perp \chi_\perp + C_{XX}\,\chi_\perp^2\big) \,,
\\ \label{q-general}
q^\mu &= 2\lambda\, \left[
	\left(C_{\phi\phi}\,\varphi_\perp 	+\tfrac{1}{2}C_{\phi X}\chi_\perp\right) \,\varphi^\mu
	+\left(\tfrac{1}{2}C_{\phi{X}}\varphi_\perp +C_{XX}\,\chi_\perp \right)\,\chi^\mu
	\right] \,,
\\ \label{tau-general}
\tau^{\mu\nu} &=-2\lambda \left(
	C_{\phi\phi}\, \varphi^\mu\varphi^\nu
	+C_{\phi{X}}\, \varphi^{(\mu}\chi^{\nu)} +C_{XX}\,\chi^\mu\chi^\nu\right) \,.
\end{align}
Taking the trace, we find the pressure 
\begin{align}\label{p-general}
p &= -\frac{2\lambda}{3}
\big(C_{\phi\phi}\,\varphi_\alpha \varphi^\alpha
+ C_{\phi X}\,\varphi_\alpha\chi^\alpha
+ C_{XX}\,\chi_\alpha \chi^\alpha
\big) \,.
\end{align}

The expressions for the pressure, energy flux, and anisotropic stress depend on the spatial vectors $\varphi^\mu$ and $\chi^\mu$. By homogeneity and isotropy these vanish on an FLRW background, implying $p{|}_{\rm FLRW}=0$, $q^\mu{|}_{\rm FLRW}=0$, $\pi^{\mu\nu}{|}_{\rm FLRW}=0$. Hence, in the cosmological background, the energy–momentum tensor \eqref{Eq:EMT} reduces to the dust form 
\begin{align}\label{Eq:EMT-dust}
T_{\mu\nu}\big{|}_{\rm FLRW} = \rho\, u_\mu u_\nu \,,
\end{align}
where $\rho=u^\alpha{u}^\beta{T}_{\alpha\beta}$ can be found for a given observer with four-velocity $u^\mu$. In this respect, our setup provide dark matter in cosmological homogeneous and isotropic background. By contrast, for standard dust \eqref{dust-EMT} one has $p=0$, $q^\mu=0$, and $\pi^{\mu\nu}=0$ identically, independent of the background or perturbations. In our setup these quantities vanish only on the homogeneous background and, as we will show in the following subsection, are generically nonzero once inhomogeneities are present. Thus the model mimics pressureless dark matter at the background level while allowing nontrivial perturbations.

The fluid four-velocity $u^\mu$ is not unique: any timelike unit vector may be used to decompose $T_{\mu\nu}$. Below we analyze two natural choices: (i) the comoving (scalar) frame, with $u_\mu \propto \partial_\mu \phi$, and (ii) the Landau (energy) frame, where $u^\mu$ is the timelike eigenvector of the energy-momentum tensor, $T^\mu{}_\alpha u^\alpha=-\rho\,u^\mu$.

\subsection{Comoving frame}
Choosing the four-velocity $u^\mu$ parallel to $\phi^\mu$ makes the comparison with the standard (no higher-derivative) case easy. Normalizing four-vector to unity gives
\begin{align}\label{u-comoving}
u^\mu={\cal N}_{\phi} \, \phi^\mu \,;
\qquad
{\cal N}_{\phi} \equiv \frac{1}{\sqrt{-X}} = \sqrt{-f}\,,
\end{align}
where in the second equality we used the generalized mimetic constraint \eqref{eq:mimetic-constraint-HD}, i.e., $X f=1$, and we choose the time-like branch $X<0$ (hence $f<0$). In this comoving/scalar frame one has $u_\mu\propto \partial_\mu\phi$, so the flow is hypersurface-orthogonal similar to the original mimetic scenario. However, $u^\mu\neq\partial^\mu\phi$ since $X\neq-1$ and therefore it does not satisfy the geodesic equation.

The acceleration in the comoving frame is given by
\begin{equation}
	a_\mu \equiv u^\alpha\nabla_\alpha u_\mu 
	= -\frac{1}{2X} V_\mu \,,
\end{equation}
where $V_\mu$ is the transverse vector defined in \eqref{V-def}. Squaring gives
\begin{align}\label{eq:C_vs_a2}
	a_\mu a^\mu
	&=\frac{1}{4X^2}\,V_\mu V^\mu =\frac{X}{4}\,{\cal C}.
\end{align}
Note that the above relation is fully covariant and holds in comoving frame. For a generic choice of $u^\mu$ not proportional to $\phi^\mu$, there is no universal identity relating $a_\mu a^\mu$ to ${\cal C}$.

Decompose the four-velocity \eqref{u-comoving} by using the relations \eqref{phi-X-decomposition}, we find
\begin{align}\label{phi-X-dec-comoving}
\varphi_\perp=-\sqrt{-X} \,, 
\qquad
\varphi^\mu=0 \,,
\qquad
\chi_\perp=\tfrac{Y}{\sqrt{-X}} \,,
\qquad
\chi^\mu
= V^\mu = - 2 X a^\mu \,,
\end{align}
which after substituting in relations \eqref{rho-general}, \eqref{q-general}, \eqref{tau-general}, and \eqref{p-general} give
\begin{align}\label{EMT-quantities-comoving-0}
	\rho &= 2\lambda\!\left( C_{\phi\phi}X + C_{\phi X}Y
	+ C_{XX}\tfrac{Y^2}{X}\right) \,,\qquad
	q^\mu = 2\lambda\,X\sqrt{-X}\left(
	C_{\phi X}
	+ 2 C_{XX}\tfrac{Y}{X}\right)a^\mu \,,\\[3pt]
	\tau^{\mu\nu} &= -8\lambda\,C_{XX}\,X^{2}\,a^\mu a^\nu ,
	\quad
	p = -\tfrac{8\lambda}{3}\,C_{XX}\,X^{2}\,a_\alpha a^\alpha ,
	\quad
	\pi^{\mu\nu}
	= -8\lambda\,C_{XX}\,X^{2}\left(a^\mu a^\nu-\tfrac{1}{3}a_\alpha a^\alpha\,h^{\mu\nu}\right). \nonumber
\end{align}
This frame makes the higher-derivative effects manifest: since $\varphi^\mu=0$, any nonzero pressure, energy flux, and anisotropic stress arise entirely from $a^\mu$, the spatial projection of $X^\mu$. Note that $X_\mu=2\,\phi^\alpha\phi_{\alpha\mu}$ and $a^\mu$ characterizes the spatial second derivatives of the scalar field.

Looking at the subset where the higher-derivative function $f$ is only function of ${\cal C}$, defined in Eq.~\eqref{C-def}, the energy-momentum coefficients simplify as shown in Eqs.~\eqref{new-coeff-def} and \eqref{eq:Chatphiphi-final} and the hydrodynamical quantities take the following simple forms
\begin{align}\label{EMT-quantities-comoving}
\begin{split}
	\rho
	&=2\lambda X\,\widehat C_{\phi\phi},
	\qquad
	q^\mu=0,
	\qquad
	p=-\frac{8\lambda}{3}\,f_{\cal C}\,a_\alpha a^\alpha,
	\\[3pt]
	\pi^{\mu\nu}
	&=-8\lambda f_{\cal C}\left(a^\mu a^\nu-\frac{1}{3}a_\alpha a^\alpha\,h^{\mu\nu}\right) \,.
\end{split}
\end{align}
We see that the energy flux vanishes in this case. This happens because, in the restricted subclass considered here, the higher-derivative corrections are arranged to probe only spatial directions (indeed, $a^\mu$ encodes only the spatial higher-derivative structure). This simplification does not hold in the general theory, where a nonzero energy flux is typically present. Therefore, for this subclass the comoving frame coincides with the Landau (energy) frame, defined by $q^\mu$=0, which we discuss in the next subsection.

Having derived the results \eqref{EMT-quantities-comoving-0} and \eqref{EMT-quantities-comoving} in the comoving frame, one can already infer a useful qualitative consequence for linear cosmological perturbations. The key point is that the acceleration $a_\mu\equiv u^\alpha\nabla_\alpha u_\mu$ is purely spatial, $u^\mu a_\mu=0$, and therefore must vanish on an exact FLRW background by homogeneity and isotropy. This is particularly transparent in unitary gauge $\phi=t$, where $a_\mu$ has only spatial components ($a_\mu\propto \delta_\mu^i\partial_i N$) and hence vanishes at the background level. For the subclass Eq.~\eqref{EMT-quantities-comoving}, all imperfect pieces induced by the higher-derivative sector are at least quadratic in $a_\mu$: the effective pressure and anisotropic stress satisfy $p\propto a_\alpha a^\alpha$ and $\pi_{\mu\nu}\propto a_\mu a_\nu-\frac{1}{3}(a_\alpha a^\alpha)h_{\mu\nu}$ (and in this subclass there is no heat flux). It follows that $\delta p$ and $\delta\pi_{\mu\nu}$ vanish at linear order around FLRW, so the higher-derivative corrections do not affect the stress tensor at the level of linear perturbations; deviations from dust-like behavior arise only beyond linear order (while the energy density may still receive corrections through the modified constraint).

\subsection{Landau frame}
In general, $q^\mu\neq 0$ in our setup. In the comoving frame we found $q^\mu \propto \big(C_{\phi X}\,X + 2 C_{XX}\,Y\big)\,V^\mu$, so it vanishes only if $V^\mu=0$ (for instance on an FLRW background) or if the prefactor happens to vanish, as in the subclass $f({\cal C})$ with ${\cal C}$ defined in Eq.~\eqref{C-def}. Alternatively, one may choose the Landau (energy) frame, defined by setting $q^\mu=0$ by construction. In this frame the four-velocity is the timelike eigenvector of $T^\mu{}_\nu$,
\begin{align}\label{Eq:eigenvalue-LL}
T^\mu{}_\alpha\,u^\alpha = -\rho\,u^\mu \,.
\end{align}

In Appendix~\ref{app:EMT-eigenvalue} we solve the eigenvalue problem for the energy–momentum tensor \eqref{Eq:EMT}. We show that, in the Landau frame, $T_{\mu\nu}$ takes the simple form
\begin{align}
T_{\mu\nu}=\rho\,u_\mu u_\nu + 3p\,n_\mu n_\nu \,,
\end{align}
where
\begin{align}\label{rho-p-def-Landau}
\rho=\lambda\!\left(T+\sqrt{T^2-4\Delta}\right),\qquad
p=-\frac{\lambda}{3}\!\left(T-\sqrt{T^2-4\Delta}\right),
\end{align}
with 
\begin{align}\label{Eq:Tr-det}
T &\equiv C_{\phi\phi}X+C_{\phi X}Y+C_{XX}Z \,,
\qquad
\Delta = \left(C_{\phi\phi}C_{XX}-\tfrac{1}{4}C_{\phi X}^2\right)
\big(XZ-Y^2\big) \,.
\end{align}
The timelike eigenvector $u^\mu$ and the spacelike vector $n^\mu$ can be written as normalized linear combinations of $\phi^\mu$ and $X^\mu$,
\begin{align}\label{eq:u-n}
u^\mu = {\cal N}_-\, \left( \phi^\mu + r_- X^\mu \right) \,,
\qquad
n^\mu = {\cal N}_+\, \left( \phi^\mu + r_+ X^\mu \right) \,,
\end{align}
satisfying
\begin{align}
u^\alpha {u}_\alpha =-1 \,,
\qquad
n^\alpha {n}_\alpha =+1 \,,
\qquad
u^\alpha n_\alpha = 0 \,.
\end{align}
The normalization factors and mixing coefficients are
\begin{align}\label{def-r-N}
{\cal N}_\pm = \frac{1}{\sqrt{\pm\left( X + 2 r_\pm Y + r_\pm^2 Z \right)}} \,,
\qquad
r_\pm = -\frac{C_{\phi\phi}X+\frac{1}{2}C_{\phi X}Y+\lambda_\pm/(2\lambda)\,}
{\,C_{\phi\phi}Y+\frac{1}{2}C_{\phi X}Z} \,,
\end{align}
where $\lambda_+=3p$ and $\lambda_-=-\rho$ are the eigenvalues in the Landau frame.

From Eq.~\eqref{rho-p-def-Landau} we see that if $\Delta=0$ then $p=0$ and the rank of $T^\mu{}_\nu$ drops from $2$ to $1$, i.e., the energy-momentum tensor reduces to that of dust. This occurs in two distinct situations. First, if $X_\mu=\eta\,\phi_\mu$ (for some scalar $\eta$), then by the definitions \eqref{XYZ-def} one has $XZ-Y^2=0$, and $T_{\mu\nu}=-2\lambda\!\left(C_{\phi\phi}+\eta\,C_{\phi X}+\eta^2 C_{XX}\right)\phi_\mu\phi_\nu$
which is rank one, i.e., proportional to the square of the single timelike one-form that is proportional to $\phi_\mu$. Secondly, if $C_{\phi\phi}C_{XX}-\tfrac{1}{4}C_{\phi X}^2=0$, $T_{\mu\nu}$ is proportional to the squared of the single timelike 1-form $\phi_\mu+\sqrt{\tfrac{C_{XX}}{C_{\phi\phi}}}X_\mu$. The usual Schutz (dust) case with action \eqref{action-dust} is a special subset of the latter case with $C_{\phi X}=C_{XX}=0$, yielding $T_{\mu\nu}=2\lambda\,\phi_\mu\phi_\nu$ and 
$\rho=2\lambda\,C_{\phi\phi}X=2\lambda$ where we used $C_{\phi\phi}=f=1/X$ (so that $Xf=1$).

\section{Avoiding caustics}\label{sec-caustic}
In this section we briefly review the standard focusing of geodesic dust, and then show that in the higher-derivative model with energy-momentum tensor \eqref{Eq:EMT} the additional terms generically induce non-geodesic acceleration and can source vorticity, thereby regulating the flow and helping to prevent caustic formation.

To analyze a congruence with tangent $u^\mu$, we decompose its covariant derivative as
\begin{equation}\label{eq:nabla-u}
\nabla_\mu u_\nu
= \frac{1}{3}\,\theta\,h_{\mu\nu}
+ \sigma_{\mu\nu}
+ \omega_{\mu\nu}
- u_\mu a_\nu \,,
\end{equation}
where
\begin{align}\label{expansion-acceleration-def}
\theta \equiv \nabla_\alpha u^\alpha \,,\qquad
a^\mu \equiv u^\alpha \nabla_\alpha u^\mu \,,
\end{align}
are the expansion scalar and the acceleration. The shear and vorticity (twist) are defined as
\begin{align}\label{shear-vorticity-def}
\sigma_{\mu\nu}
\equiv h_\mu{}^{\alpha} h_\nu{}^{\beta}
\left( \nabla_{(\alpha} u_{\beta)} - \tfrac{1}{3}\,\theta\, h_{\alpha\beta} \right) \,,
\qquad
\omega_{\mu\nu}
\equiv h_\mu{}^{\alpha} h_\nu{}^{\beta}
\nabla_{[\alpha} u_{\beta]} \,.
\end{align}
By construction, shear and vorticity are spatial tensors
\begin{align}
u^\alpha \sigma_{\alpha\mu} = 0 \,,
\qquad
u^\alpha \omega_{\alpha\mu} = 0 \,,
\end{align}
and $\omega_{\mu\nu}=0$ if $u_\mu$ is hypersurface-orthogonal.

The $3\!+\!1$ decomposition defines derivatives along and orthogonal to the flow as
\begin{align}\label{derivatives-def}
\frac{{\rm D}}{\D\tau} \equiv u^\alpha \nabla_\alpha \,,
\qquad
D_\mu \equiv h_\mu{}^{\alpha}\nabla_\alpha \,,
\end{align}
where $\tau$ is the proper time along the congruence. Taking the derivative along $u^\mu$ of Eq.~\eqref{eq:nabla-u} and tracing yields the Raychaudhuri equation
\begin{equation}\label{eq:Raychaudhuri}
\dot\theta
= -\frac{1}{3}\theta^{2}
-\sigma_{\alpha\beta}\sigma^{\alpha\beta}
+\omega_{\alpha\beta}\omega^{\alpha\beta}
- R_{\alpha\beta}u^\alpha u^\beta
+ \nabla_\alpha a^\alpha \,,
\end{equation}
where a dot denotes ${\rm D}/\D\tau$. Since $\sigma_{\mu\nu}$ and $\omega_{\mu\nu}$ are spatial (orthogonal to $u^\mu$), their quadratic contractions are non-negative
\begin{align}\label{eq:spatial-sigma-omega}
\sigma_{\alpha\beta}\sigma^{\alpha\beta}\ge 0 \,,
\qquad
\omega_{\alpha\beta}\omega^{\alpha\beta}\ge 0 \,,
\end{align}
with equality holding if the corresponding tensor vanishes.

The sign of the Ricci term in the Raychaudhuri equation \eqref{eq:Raychaudhuri} follows from Einstein’s equations
\begin{align}\label{Ruu}
\Mpl^2 R_{\alpha\beta} u^\alpha u^\beta = T_{\alpha\beta} u^\alpha u^\beta + \tfrac{1}{2} T^{\alpha}{}_\alpha = \tfrac{1}{2} \left(\rho+3p\right) \,.
\end{align}
If the strong energy condition holds, then in particular $\rho+3p\ge0$, implying $R_{\alpha\beta}u^\alpha u^\beta\ge0$. One may allow for $\rho+3p<0$, which violates the strong energy condition and yields a repulsive contribution. This is possible for the scalar sector since $\rho$ and $p$ refer to the effective fluid built from $\phi$. Including additional matter, $\Mpl^2 R_{\alpha\beta} u^\alpha u^\beta = \left[\rho+\rho_\m+3(p+p_m)\right]/2$. Even if the matter sector alone satisfies the strong energy condition, $\rho_m+3p_m\ge0$, the total can still violate it if the scalar contribution dominates. In what follows we adopt the conservative assumption that the total stress–energy satisfies the strong energy condition, so that $R_{\alpha\beta}u^\alpha u^\beta\ge0$.

\subsection{Caustics in dust}
Before analyzing our higher-derivative setup, let us first apply the above formalism to the case of standard pressureless dust. This brief review of caustic formation in dust (see Ref.~\cite{Wald:1984rg}) will help for our discussion in the next subsection.

The standard dust case can be recovered from \eqref{Eq:EMT} by setting $C_{\phi X}=C_{XX}=0$ and $C_{\phi\phi}=-1$ where we used the dust constraint $X=-1$. Then
\begin{align}
T_{\mu\nu}=\rho\,u_\mu u_\nu \,,
\qquad
\rho=2\lambda \,,
\qquad
u_\mu=\phi_\mu \,.
\end{align}
Using $X=-1$, we immediately find $a^\mu\equiv u^\alpha\nabla_\alpha u^\mu=0$ and since the flow is hypersurface-orthogonal, $\omega_{\mu\nu}=0$. Hence the Raychaudhuri equation \eqref{eq:Raychaudhuri} simplifies to
\begin{equation}
\dot\theta = -\frac{1}{3}\theta^2-\sigma_{\alpha\beta}\sigma^{\alpha\beta} - \frac{\rho}{2\Mpl^2} \,;
\qquad
\theta = \Box\phi \,,
\end{equation}
where we have used \eqref{Ruu}, imposing $p=0$ for the dust. Using \eqref{eq:spatial-sigma-omega}, and the fact that $\rho>0$ (even weak energy condition is enough in the case of dust), we find
\begin{align}\label{Eq:theta}
\frac{{\rm D}}{\D\tau}\left(\frac{1}{\theta}\right) \ge \frac{1}{3}
\quad
\Longrightarrow
\quad
\frac{1}{\theta(\tau)} \;\ge\; \frac{1}{\theta(\tau_0)} + \frac{1}{3}\,(\tau - \tau_0) \,,
\end{align}
where $\tau_0$ is some initial proper time. 

Suppose that $\theta(\tau_0) < 0$. But the right hand side of \eqref{Eq:theta} increases linearly with $\tau$. Thus, starting from a negative value, after some finite time it will pass through zero. That means $1/\theta$ goes from negative to zero in finite time so that $\theta \to -\infty$. The first zero of the right-hand side of \eqref{Eq:theta} gives an upper bound on the blow-up time $\tau_c$
\begin{align}\label{tau-c}
\tau_c \equiv \tau_0 + \tfrac{3}{|\theta(\tau_0)|} \,,
\end{align}
such that $\theta(\tau)\to -\infty$ for some $\tau\le \tau_c$.

The continuity equation $\nabla_\alpha{T}^\alpha{}_\mu=0$ implies
\begin{align}
\dot{\rho} + \theta \rho = 0
\quad\Longrightarrow\quad
\rho(\tau)=\rho_0\,\exp\!\left(-\int_{\tau_0}^{\tau}\!\theta(\tilde\tau)\,d\tilde\tau\right) \,,
\end{align}
which shows that when $\theta(\tau)\to -\infty$, energy density diverges.

The above results imply that, along the geodesic, irrotational dust flow,
\begin{align}\nonumber
\rho=2\lambda \;\to\; \infty \,,
\qquad
\Box\phi=\nabla_\mu u^\mu \;\to\; -\infty \,,
\end{align}
signalling a caustic (shell-crossing) singularity when neighboring geodesics intersect. In a purely hydrodynamic interpretation, where $\phi$ is a velocity potential, such caustics indicate the breakdown of the fluid description rather than a fundamental pathology. In mimetic gravity, however, $\phi$ encodes the conformal mode of the metric, so the onset of caustics corresponds to a singular mimetic field and thus to a breakdown of the gravitational construction itself.

\subsection{Higher derivatives: non-vanishing acceleration and vorticity}
In the dust case \eqref{dust-EMT}, the acceleration and vorticity vanish, $a^\mu=0$ and $\omega_{\mu\nu}=0$, so the Raychaudhuri equation \eqref{eq:Raychaudhuri} integrates to a finite focusing time \eqref{tau-c} for the formation of caustic singularities. In contrast, for the higher-derivative energy-momentum tensor \eqref{Eq:EMT} we show that both $a^\mu$ and $\omega_{\mu\nu}$ are generically nonzero.

The divergence of the acceleration appearing in the Raychaudhuri equation \eqref{eq:Raychaudhuri} can be written as
\begin{align}\label{a-div}
\nabla_\alpha a^\alpha = D_\alpha a^\alpha + a^\alpha a_\alpha \,,
\end{align}
where $D_\mu$ is the spatial covariant derivative defined in \eqref{derivatives-def}. Since $a^\mu$ is spatial ($u_\alpha a^\alpha=0$), one has $a^\alpha a_\alpha\ge 0$. Thus the last term in \eqref{a-div} contributes positively on the right-hand side of \eqref{eq:Raychaudhuri}, opposing geodesic focusing and helping to prevent caustic formation. The sign of the first term, $D_\alpha a^\alpha$, is not fixed in general. To get better understanding of $D_\alpha a^\alpha$, let us obtain an explicit expression for the acceleration. 
It is useful to start with the scalar comoving frame $u^\mu\propto\phi^\mu$, where one has $a_\mu=-V_\mu/(2X)$. Taking a spatial divergence gives
\begin{equation}\label{eq:Da_comoving}
	D_\alpha a^\alpha
	= -\frac{1}{2X}\,D_\alpha V^\alpha
	+\frac{1}{2X^2}\,V^\alpha D_\alpha X,
	\qquad \mbox{comoving frame}
\end{equation}
so while $a_\alpha a^\alpha\ge0$ is sign-definite, the sign of $D_\alpha a^\alpha$ depends on the local spatial profile of $X$ (and thus on inhomogeneities). 
For example, in regimes where spatial gradients are dominated by the scalar $X$ itself so that $V_\mu\simeq D_\mu X$ and $D_\alpha V^\alpha\simeq D^2X$, one has $D_\alpha a^\alpha\simeq -(2X)^{-1}D^2X+\cdots$; since $X<0$ on the timelike branch, a local condition such as $D^2X>0$ then tends to give $D_\alpha a^\alpha>0$ and hence contributes to defocusing in the Raychaudhuri equation. 
More generally, however, $D_\alpha a^\alpha$ can have either sign because both terms in \eqref{eq:Da_comoving} contribute. In the Landau frame ($q^\mu=0$), projecting $\nabla_\alpha T^{\mu\alpha}=0$ orthogonally to $u^\mu$ yields the relativistic Euler equation
\begin{equation}\label{eq:EulerLandau}
	a_\mu = \frac{F_\mu}{\rho+p} \,,
	\qquad
	F_\mu \equiv - D_\mu p - D_\alpha \pi^{\alpha}{}_{\mu} \,.
	\qquad \mbox{Landau frame}
\end{equation}
The vector $F_\mu$ is the relativistic analogue of a force density, so spatial gradients of $p$ and $\pi^{\mu\nu}$ generically imply $a_\mu\ne 0$.

If the vorticity is non-vanishing $\omega_{\mu\nu}\neq0$, its contribution to the right hand side of the Raychaudhuri equation \eqref{eq:Raychaudhuri} is always positive $+\omega_{\alpha\beta}\omega^{\alpha\beta}\geq0$ as it is a spatial tensor. However, the value of $\omega_{\mu\nu}$ depends on the frame. Substituting \eqref{u-comoving} and \eqref{eq:u-n} in the definition \eqref{shear-vorticity-def}, we find
\begin{align}
\omega_{\mu\nu} =
\begin{cases}
D_{[\mu}\!\ln\mathcal N_\phi\;\varphi_{\nu]}
= 0 \,,
&\qquad \mbox{comoving frame}
\\
D_{[\mu}\!\ln\mathcal N_{-}\,\big(\varphi_{\nu]} + r_{-}\,\chi_{\nu]}\big)
\;+\; \mathcal N_{-}\, D_{[\mu} r_{-}\ \chi_{\nu]} \,,
&\qquad \mbox{Landau frame}
\end{cases}
\end{align}
where $\mathcal N_{-}$ and $r_{-}$ are given in \eqref{def-r-N} and we have used the fact that $\varphi^\mu=0$ in the comoving frame (see Eq. \eqref{phi-X-dec-comoving}). Hence $\omega_{\mu\nu}\neq 0$ in general.

For completeness, we compute the expansion parameter in comoving and Landau frames by substituting \eqref{u-comoving} and \eqref{eq:u-n} in the definition \eqref{expansion-acceleration-def} yielding
\begin{align}\label{theta-HD}
\theta =
\begin{cases}
\frac{1}{\sqrt{-X}}\!\left(\Box\phi - \tfrac12\,\phi^\mu\nabla_\mu\ln(-X)\right) \,,
&\mbox{comoving frame}
\\
\mathcal N_{-}\,\Box\phi
+ \frac{{\rm D}}{{\rm d}\tau}\ln\mathcal N_{-}
+ \mathcal N_{-}\, r_{-}\!\left(\Box X + X^\alpha \nabla_\alpha \ln r_{-}\right) \,,
&\mbox{Landau frame}
\end{cases}
\end{align}
where we note that the proper time in the Landau frame is defined along the corresponding four-velocity given by \eqref{eq:u-n}. The comparison with the original mimetic dust scenario is very clear in the comoving frame: $X=-1$ and we find $\theta=\Box\phi$. 

In summary, for dust \eqref{dust-EMT} with $a^\mu=0$, $\omega_{\mu\nu}=0$, and $u^\alpha u^\beta R_{\alpha\beta}\!\ge\!0$, any region with $\theta<0$ inevitably reaches $\theta\to-\infty$ in finite proper time. By contrast, in our higher-derivative model \eqref{Eq:EMT}, the acceleration and the vorticity are generically nonzero. Consequently, the positive contribution $+\omega_{\alpha\beta}\omega^{\alpha\beta}$ and the non-negative piece $+a_\alpha a^\alpha\subset\nabla_\alpha a^\alpha$ in the Raychaudhuri equation \eqref{eq:Raychaudhuri} can counteract geodesic focusing and help prevent caustic formation. Therefore, the necessary condition is at least one of $a^\mu$ or $\omega_{\mu\nu}$ be nonzero while the sufficient condition is
\begin{equation}\label{eq:sufficient}
\omega_{\alpha\beta}\omega^{\alpha\beta}+\nabla_\alpha a^\alpha
\;\ge\;
\sigma_{\alpha\beta}\sigma^{\alpha\beta}
+ u^\alpha u^\beta R_{\alpha\beta}
+ \frac{1}{3}\theta^2 \,.
\end{equation}
If along every world-line \eqref{eq:sufficient} holds for all proper times, then $\dot\theta\ge 0$ and $\theta$ cannot diverge to $-\infty$ and hence no caustic forms.

It should be stressed that the above criterion is meant as a structural indication rather than an explicit proof of caustic avoidance in a given configuration: establishing whether \eqref{eq:sufficient} is actually satisfied along the flow for representative inhomogeneous solutions would require solving the coupled system (analytically in symmetry-reduced setups or numerically), which is beyond the scope of the present work. Nevertheless, an important consistency aspect of the subclass $f({\cal C})$ identified in Sec.~\ref{sec-HD-mimetic} is that, in unitary gauge, the higher-derivative contributions enter through spatial derivatives only. As a result, the evolution equations in this sector do not acquire higher time derivatives, and the initial-value formulation is not expected to be obstructed by Ostrogradsky-type pathologies, while the new spatial-gradient terms can become relevant precisely in regimes where inhomogeneities grow and the standard dust congruence would otherwise undergo geodesic focusing. This provides a concrete setting in which the mechanism suggested by \eqref{eq:sufficient} can operate, and motivates a dedicated study of explicit solutions and well-posedness properties as a separate follow-up.

The idea that higher-derivative structures can prevent caustic formation has been appeared in other modified gravity scenarios. In projectable Ho\v{r}ava-Lifshitz gravity, it is argued that higher-curvature terms can prevent formation of caustics singularities \cite{Mukohyama:2009tp}. Moreover, several works make explicit the close connection between mimetic gravity and the infrared limit of projectable Ho\v{r}ava-Lifshitz gravity, suggesting a common mechanism whereby the extra mode sources non-geodesic flow \cite{Lim:2010yk,Ramazanov:2016xhp,Babichev:2017lrx,Chamseddine:2019gjh}.
	
\section{Summary and conclusions}\label{sec:summary}
In standard $\Lambda$CDM cosmology, dark matter is modeled as pressureless dust, whose action \eqref{action-dust} is found long time ago \cite{Schutz:1977df}. More recently, within mimetic gravity \cite{Chamseddine:2013kea,Golovnev:2013jxa}, the same dust dynamics was obtained by imposing a singular conformal transformation on the Einstein-Hilbert action, thereby encoding the dust-like component in the metric’s conformal mode. Because the dust four-velocity is geodesic (and irrotational), geodesic congruences generically focus and form caustics in finite proper time. For ordinary dark matter, treated as a phenomenological fluid, such caustics signal only the breakdown of the fluid description. In the mimetic setting, however, they correspond to singular behavior of the mimetic field and hence to a breakdown of the gravitational construction itself.

Nevertheless, the mimetic construction provides a systematic approach to the dust action by looking at the singular limit of a conformal transformation of the metric in the gravitational sector. In this work we followed that approach to incorporate higher-derivative effects in a systematic way. In particular, consistency of the underlying disformal map restricts the allowed higher-derivative structures, leading to the generalized action \eqref{action}. Unlike earlier approaches, where higher-derivative operators were added directly to the original mimetic Lagrangian, here all higher-derivative corrections are encoded solely in the (generalized) mimetic constraint \eqref{mimetic-constraint}. In general, establishing the absence of Ostrogradsky-like ghost degrees of freedom for the full theory \eqref{action} requires a complete Hamiltonian analysis. On the other hand, there exists a particularly simple and well-controlled subclass, obtained by restricting to $f({\cal C})$ with ${\cal C}$ defined in \eqref{C-def}, for which the potentially dangerous higher-time-derivative structure is absent already at the level of the operator basis (and in unitary gauge the higher-derivative contributions enter through spatial derivatives only). The resulting energy-momentum tensor describes an imperfect fluid and reduces to dust when the higher-derivative terms are switched off. We showed that, on homogeneous cosmological backgrounds the theory behaves as pressureless dark matter, whereas in the presence of inhomogeneities the higher-derivative terms generate nonzero acceleration (non-geodesic flow) and vorticity, thereby helping to prevent caustic formation.
	
\subsubsection*{Acknowledgements}
 I am grateful to Karim Noui and Pavel Petrov for helpful discussions. This work was supported by IBS under the project code IBS-R018-D3.
	
\appendix

\section{Eigenproblem for the energy-momentum tensor}\label{app:EMT-eigenvalue}

In this appendix we solve the eigenproblem
\begin{align}\label{app:eq-eigen}
T^\mu{}_\alpha v^\alpha_a = \lambda_a v^\mu_a \,,
\end{align}
for the energy-momentum tensor \eqref{Eq:EMT}. Our aim is to find explicit forms of the eigenvalues $\lambda_a$ and the corresponding eigenvectors $v^\mu_a$. 

The energy-momentum tensor \eqref{Eq:EMT} has rank~2 which can be expressed in the 2-plane that is spanned by the two vectors $\{\phi^\mu,X^\mu\}$. Therefore the eigenvectors should have a general form
\begin{align}\label{app:v-gen}
v^\mu_a = {\cal N}_a\, \left( \phi^\mu + r_a X^\mu \right) \,,
\end{align}
where ${\cal N}_a$ and $r_a$ are coefficients which will be determined soon. Substituting \eqref{app:v-gen} in \eqref{app:eq-eigen}, we find
\begin{align}\label{app:eq-eigen-r}
\begin{pmatrix}
	T_{11} & T_{12} \\
	T_{21} & T_{22}
\end{pmatrix} 
\cdot
\begin{pmatrix}
	1 \\
	r_a
\end{pmatrix}
= \lambda_a \begin{pmatrix}
	1 \\
	r_a
\end{pmatrix}, 
\end{align}
where
\begin{align}
T_{IJ} = -2\lambda
\begin{pmatrix}
C_{\phi\phi}X+\tfrac{1}{2}C_{\phi X}Y &
C_{\phi\phi}Y+\tfrac{1}{2}C_{\phi X}Z\\[3pt]
\tfrac{1}{2}C_{\phi X}X+C_{XX}Y &
\tfrac{1}{2}C_{\phi X}Y+C_{XX}Z
\end{pmatrix} \,.
\end{align}
The above reduction means that $T^\mu{}_\nu$ has two zero eigenvalues and the other two nonzero eigenvalues are given by eigenvalues of $T_{IJ}$
\begin{align}
\lambda_{\pm}=\tfrac{1}{2}\,\Tr T_{IJ}
\ \pm \tfrac{1}{2}\sqrt{\big(\Tr T_{IJ}\big)^2 - 4\det(T_{IJ})} \,,
\end{align}
with
\begin{align}
\label{app:Eq:Tr-det}
\begin{split}
\Tr T_{IJ} &= -2\lambda\,(C_{\phi\phi}X+C_{\phi X}Y+C_{XX}Z) \,,
\\
\det(T_{IJ}) &= 4\lambda^2\left(C_{\phi\phi}C_{XX}-\tfrac{1}{4}C_{\phi X}^2\right)
\big(XZ-Y^2\big) \,.
\end{split}
\end{align}
Note that clearly $\Tr T_{IJ} =\Tr T^\mu{}_\nu$ since the two zero eigenvalues of $T^\mu{}_\nu$ do not contribute.

The limit $C_{\phi{X}}=0=C_{XX}$ corresponds to dust where the energy momentum is completely characterized by the four-velocity $u^\mu$ and energy density $\rho$. In this case, $\det(T_{IJ})=0$ and $\lambda_+=0$ and $\lambda_-=\Tr T_{IJ}$ since $\Tr T_{IJ}<0$, which shows that $\lambda_-$ characterizes the energy density. This implies $u^\mu \equiv v^\mu_-$
\begin{align}\label{app:u-n}
u^\mu = {\cal N}_-\, \left( \phi^\mu + r_- X^\mu \right) \,,
\qquad
n^\mu = {\cal N}_+\, \left( \phi^\mu + r_+ X^\mu \right) \,,
\end{align}
where $n^\mu \equiv v^\mu_+$ is another spacelike four-vector. Imposing normalization conditions
\begin{align}
u^\alpha {u}_\alpha =-1 \,,
\qquad
n^\alpha {n}_\alpha =+1 \,,
\end{align}
give
\begin{align}
{\cal N}_\pm = \frac{1}{\sqrt{\pm\left( X + 2 r_\pm Y + r_\pm^2 Z \right)}} \,.
\end{align}
Note that $u^\alpha n_\alpha = 0$. The explicit values of $r_{\pm}$ can be found from \eqref{app:eq-eigen-r} as
\begin{align}
r_\pm = -\frac{C_{\phi\phi}X+\frac{1}{2}C_{\phi X}Y+\lambda_\pm/(2\lambda)\,}
{\,C_{\phi\phi}Y+\frac{1}{2}C_{\phi X}Z} \,.
\end{align}
Substituting $\phi^\mu$ and $X^\mu$ from \eqref{app:u-n} in \eqref{Eq:EMT}, we find
\begin{align}
T_{\mu\nu} = - \lambda_- u_\mu u_\nu + \lambda_+ n_\mu n_\nu \,.
\end{align}
Comparing the above result with \eqref{EMT-dec} while $q^\mu=0$, we find 
\begin{align}
\rho=-\lambda_- \,,
\qquad
p = \tfrac{1}{3} \lambda_+ \,,
\qquad
\tau^{\mu\nu} = \lambda_+ n^\mu n^\nu \,,
\qquad
\pi^{\mu\nu} = \lambda_+ \left(n^\mu n^\nu - \tfrac{1}{3} h^{\mu\nu}\right)\,. 
\end{align}

\bibliographystyle{JHEPmod}
\bibliography{refs}
	
\end{document}